\documentclass[apjl,iop,numberedappendix,twocolappendix]{emulateapj}

\usepackage{mathtools}
\usepackage{esint,graphicx,hyperref}
\hypersetup{colorlinks=false,pdfborder={0 0 0}}
\setcitestyle{notesep={; }}
\usepackage[normalem]{ulem}

\usepackage{graphicx}	
\usepackage{amsmath}	
\usepackage{amssymb}	
\usepackage{color}

\defcitealias{EHT1}{EHT1}
\defcitealias{EHT2}{EHT2}
\defcitealias{EHT3}{EHT3}
\defcitealias{EHT4}{EHT4}
\defcitealias{EHT5}{EHT5}
\defcitealias{EHT6}{EHT6}

\slugcomment{Accepted for publication in ApJL, October 25, 2019}

\begin{document}

\title{The Shadow of a Spherically Accreting Black Hole}
\shorttitle{Black Hole Shadow}

\author{Ramesh~Narayan\altaffilmark{1}, Michael~D.~Johnson\altaffilmark{1}, and Charles~F.~Gammie\altaffilmark{2,3}}
\shortauthors{Narayan et al.}
\altaffiltext{1}{Center for Astrophysics $|$ Harvard \& Smithsonian, 60 Garden Street, Cambridge, MA 02138, USA}
\altaffiltext{2}{Department of Physics, University of Illinois, 1110 West Green St, Urbana, IL 61801, USA}
\altaffiltext{3}{Department of Astronomy, University of Illinois at Urbana-Champaign, 1002 West Green Street, Urbana, IL 61801, USA}
\email{rnarayan@cfa.harvard.edu} 


\begin{abstract}
We explore a simple spherical model of optically thin accretion on a
Schwarzschild black hole, and study the properties of the image as
seen by a distant observer. We show that a dark circular region in the
center --- a shadow --- is always present. The outer edge of the
shadow is located at the photon ring radius $b_{\rm ph} \equiv
\sqrt{27}r_g$, where $r_g=GM/c^2$ is the gravitational radius of the
accreting mass $M$.  The location of the shadow edge is independent of
the inner radius at which the accreting gas stops radiating. The size
of the observed shadow is thus a signature of the spacetime geometry
and it is hardly influenced by accretion details.  We briefly discuss
the relevance of these results for the Event Horizon Telescope image
of the supermassive black hole in M87.

\end{abstract}

\section{Introduction}

Recently, the Event Horizon Telescope (EHT) obtained an ultra high
angular resolution image of the accretion flow around the supermassive
black hole in M87 (\citealt{EHT1,EHT2,EHT3,EHT4,EHT5,EHT6}; hereafter
EHT1-6). The image shows a bright ring of emission surrounding a dark
interior \citepalias{EHT4}. The presence of such a dark central region was
predicted by \citet{falcke}, who termed it the ``shadow of the black
hole.''

The shadow phenomenon is caused by gravitational light deflection --
gravitational lensing -- by the black hole. Lensing by a black hole
was studied earlier in the context of an extended background source
\citep{bardeen} and for a geometrically thin, optically thick
accretion disk \citep{luminet}. When the accretion flow produces
radiation all around the black hole and the gas is optically thin to
its own radiation, \citet{falcke} \citep[see also][]{jaroszynski} showed that lensing still produces a
dramatic signature in the form of a shadow. They noted that the effect
should be visible in Sagittarius A$^*$, the supermassive black hole at
the center of our Galaxy, which is known to have a hot optically-thin
accretion flow. Similar hot accretion flows are found around many
supermassive black\ holes in the universe \citep{yuan}, including the
black hole in M87. All of these objects are natural candidates to
reveal shadows in their images. The discovery by the EHT of a
shadow-like feature in the image of M87 is thus 
verification of a fundamental prediction of general relativity in the regime of
strong lensing.

As described in many previous studies, the outer edge of the shadow in
a black hole image is located at the ``photon ring,'' which is the
locus of rays that escape from bound photon orbits around the black
hole to a distant observer. The photon ring, as defined here, was
called the ``apparent boundary'' by \citet{bardeen} and the ``critical
curve'' by \citet{gralla}. The latter authors questioned whether the
EHT did in fact see a shadow surrounded by the photon ring in M87;
they presented some models in which the details of the accretion flow
seemed to matter more than any particular relativistic phenomenon such
as the photon ring.

In this paper we use a simple spherical model of black hole accretion
to study the properties of the shadow and the photon ring. After some
preliminaries in \S2, we discuss in \S3 a Newtonian model to clarify
the notion of the shadow of an optically thin accretion flow in flat
space. We then consider in \S4 and \S5 two general relativistic models
that use the Schwarzschild spacetime. We show that the shadow is a
robust feature of these latter models and that its size and shape are
primarily influenced by the spacetime geometry, and not by details of
the accretion.  We conclude in \S6 with a discussion.

\section{Terminology and Notation}

We consider spherical accretion on a gravitating object of mass $M$.
We use $r_g = GM/c^2$ as our unit of length and $c^2$ as our unit of
energy. In the case of the Newtonian model (\S3), we take the radius
of the central mass to be $2r_g$ and assume flat spacetime exterior to
the surface. We also assume that any light rays that fall on the
surface are completely absorbed.

For the general relativistic model (\S4, \S5), we consider a
Schwarzschild black hole with spacetime metric
\begin{equation}
ds^2 = g_{tt} dt^2 + g_{rr} dr^2 + r^2 d\theta^2 + r^2 \sin^2\theta\; d\phi^2,
\end{equation}
where $t$ is time, $r$ is radius, and
\begin{equation}
g_{tt} = -\left(1-\frac{2}{r}\right), \quad
g_{rr} = \left(1-\frac{2}{r}\right)^{-1}.
\end{equation}
The horizon of the black hole is at $r=2$, and the spacetime has an
unstable circular ``photon orbit'' at radius
\begin{equation}
r_{\rm ph} = 3.
\end{equation}
A finely tuned photon at the photon orbit could, in principle, orbit
the black hole an infinite number of times. However, since the orbit
is unstable, any slight perturbation would cause the photon either to
fall into the black hole or to escape to infinity. The photons that
escape are seen by a distant observer to have an impact parameter
$b_{\rm ph}$ with respect to the mass $M$, where
\begin{equation}
b_{\rm ph}=\sqrt{27} \approx 5.2.
\end{equation}
We refer to the circle in the observer's sky with radius $b_{\rm ph}$,
centered on $M$, as the ``photon ring.''

\section{Newtonian Model}\label{sec:flat}

In Newtonian gravity, the potential energy of accreting gas varies
with radius $r$ as $-1/r$. Hence, we consider the following physically
reasonable model of the power per unit volume $P_{\rm N}$ (${\rm
  erg\,cm^{-3}s^{-1}}$) radiated by the gas,
\begin{equation}
P_{\rm N}(r) = \frac{1}{4 \pi r^4}, \quad r\geq2.
\end{equation}
This is bolometric power, integrated over frequency.  The total
luminosity $L_{\rm emit}(r)$ emitted between radius $r$ and infinity
is
\begin{equation}
L_{\rm emit}(r) = \int_r^\infty P_{\rm N}(r')\; 4 \pi r'^2 dr' = \frac{1}{r},
\end{equation}
i.e., proportional to the total potential energy drop from infinity to $r$, as
is reasonable for accretion.  The luminosity of an accretion flow
depends of course on the mass accretion rate, so we ought to multiply
$P_{\rm N}$ by a proportionality constant. However, under the fully
optically thin conditions we consider in this paper, none of the
results depend on the constant, so for simplicity we ignore it.  The
emission coefficient per unit solid angle $j_{\rm N}$ (${\rm
  erg\,cm^{-3}s^{-1}ster^{-1}}$) in this model is
\begin{equation}
j_{\rm N}(r) = \frac{P_{\rm N}(r)}{4\pi {\rm (ster)}} = \frac{1}{16 \pi^2 r^4}, \quad r\geq2.
\end{equation}

Since the gas is optically thin, rays can travel arbitrarily large
distances without being absorbed or scattered. However, not all of the
emitted radiation escapes to infinity; any ray that intersects the
``surface'' of the central object at $r=2$ is absorbed and lost. In
the Newtonian approximation considered in this section (flat space),
rays travel in straight lines. Hence for radiation emitted at radius
$r$, the solid angle corresponding to rays that are lost is
$2\pi(1-\cos\theta)$, while that of rays that escape to infinity is
$2\pi(1+\cos\theta)$, where $\sin\theta=2/r$.  Thus the net luminosity
observed at infinity is
\begin{align}
L_\infty &= \int_2^\infty j_{\rm N}(r)\; 2\pi\left[1 + \sqrt{1-\frac{4}{r^2}}
\right]4\pi r^2dr\\ 
\nonumber &= \frac{1}{4}+\frac{\pi}{16} \approx 0.45.
\end{align}

We are interested in the image that a distant observer, located say on
the $z$-axis, would see. For this, we compute the observed intensity
$I(b)$ (${\rm erg\,cm^{-2}s^{-1}ster^{-1}}$) as a function of impact
parameter $b$.  For $b\geq2$, this is given by
\begin{equation}
I(b) = \int_{-\infty}^\infty j_{\rm N}\left(\sqrt{b^2+z^2}\right)dz =
\frac{1}{32\pi b^3}, \quad b\geq2,
\end{equation}
while for smaller impact parameters it is given by
\begin{eqnarray}
  I(b) &=& \int_{\sqrt{4-b^2}}^\infty j_{\rm N}\left(\sqrt{b^2+z^2}\right)dz  \\
\nonumber &=&
\frac{1}{32\pi^2 b^3}\tan^{-1}\left(\frac{b}{\sqrt{4-b^2}} \right)
-\frac{\sqrt{4-b^2}}{128\pi^2 b^2},
~~ b<2.
\end{eqnarray}

\begin{figure}
\includegraphics[width=\columnwidth]{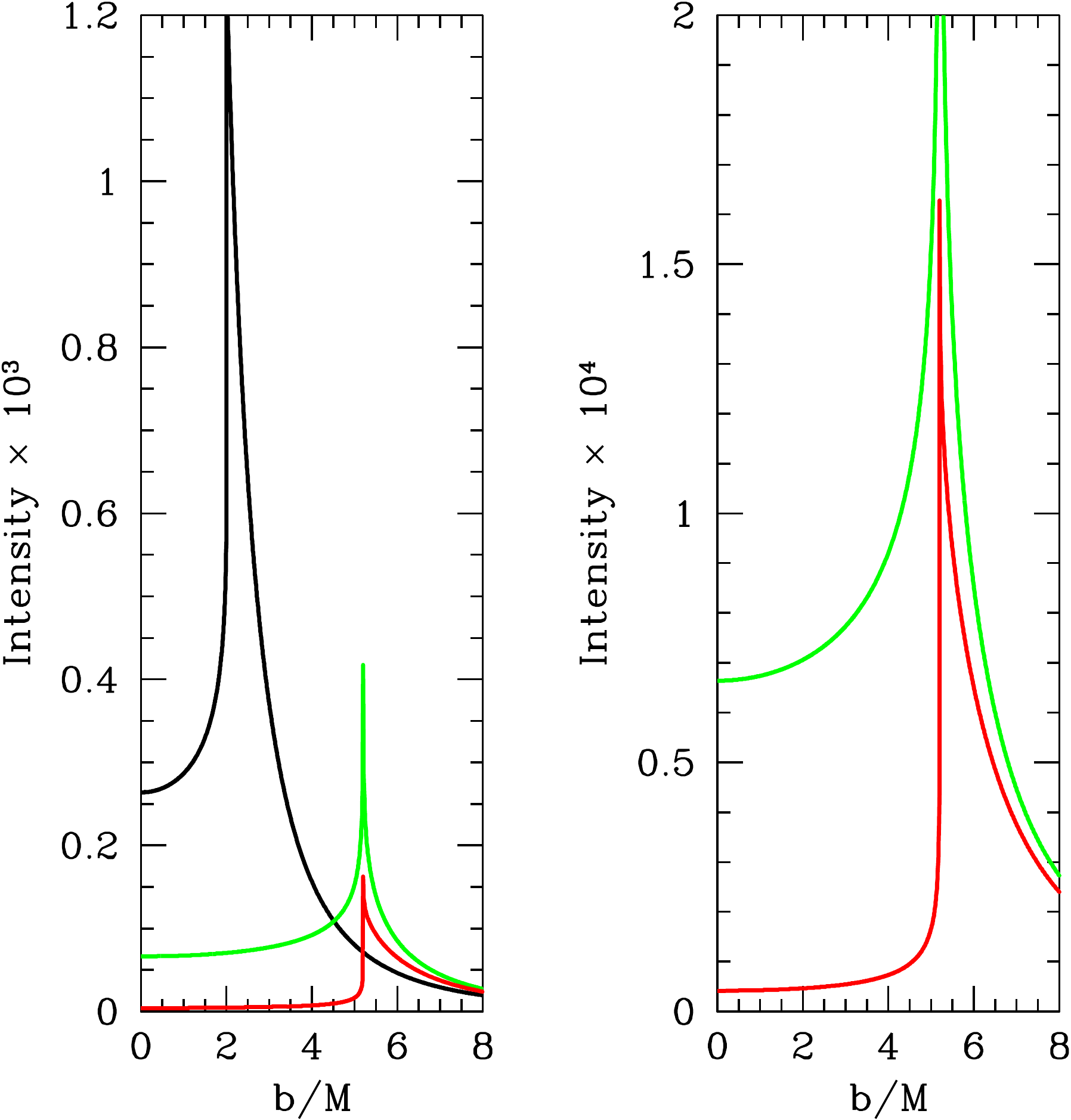}
\caption{Left: Shows the intensity profile $I(b)$ as a function of the
  impact parameter $b$ as seen by a distant observer for a spherically
  symmetric problem. The curves correspond to the Newtonian model
  (black), Schwarschild spacetime with radiating gas at rest (green)
  and Schwarzschild spacetime with radially free-falling gas
  (red). Right: The two Schwarzschild models shown on an expanded
  vertical scale.}
\label{fig:intensity_profiles}
\end{figure}

\begin{figure*}
\includegraphics[width=\textwidth]{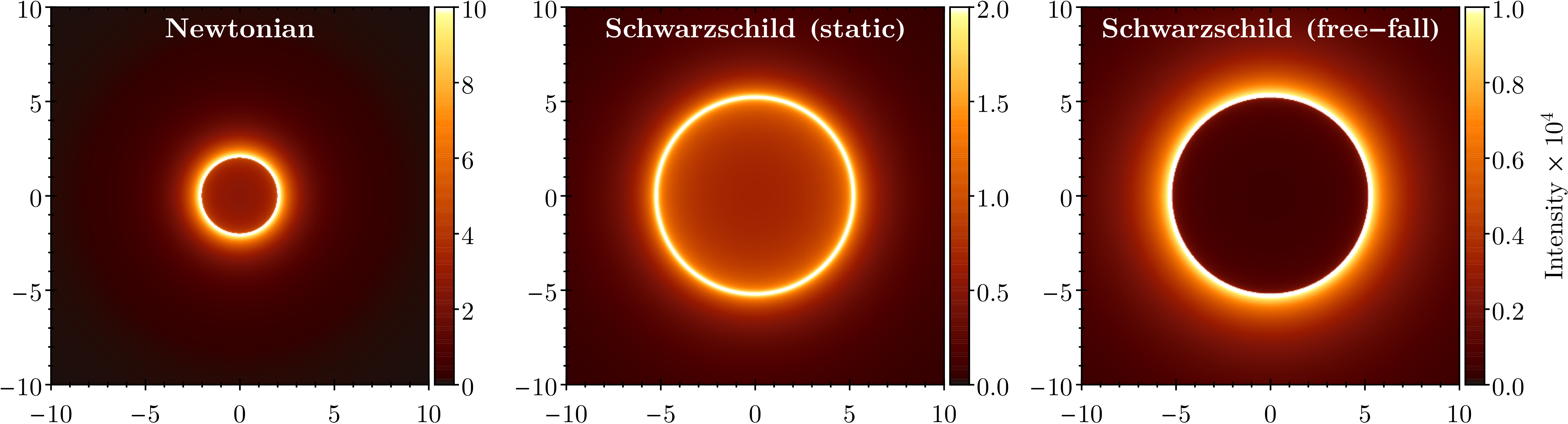}
\caption{Image seen by a distant observer for the same models as in
  Fig.~\ref{fig:intensity_profiles}: Newtonian model (left), Schwarzschild spacetime with gas at rest (middle), and Schwarzschild spacetime with radially
  infalling gas (right).}
\label{fig:2D}
\end{figure*}

The black curve in Fig.~\ref{fig:intensity_profiles} shows the radial
intensity profile $I(b)$ for this Newtonian model, and the left panel
in Fig.~\ref{fig:2D} shows the two-dimensional image seen by a distant
observer. The intensity increases monotonically with decreasing impact
parameter as $1/b^3$, until it reaches a maximum value equal to
$1/256\pi$ at the edge of the central object ($b=2$).  Inside this
critical impact parameter, the intensity drops suddenly by a factor of
2 since the line-of-sight terminates on the surface of the central
mass and one no longer sees emission from the far side. It then
continues to decrease inward and reaches a value of $1/384\pi^2 \approx 2.64\times 10^{-4}$ at
$b=0$. The intensity at the center is 4.7 times lower than the peak
intensity at $b=2$.

The dark region inside $b=2$ is what we term the ``shadow'' in the
image.  It is not a totally dark shadow with zero intensity, such as
would be observed if the radiating gas were entirely behind the
mass. Since we have radiating gas all around the mass, the image has
some intensity even for lines-of-sight that intersect the surface of
the central mass. Nevertheless, the shadow is a distinct feature. Its
radius is well-defined and could be measured by a distant observer
with sufficiently good angular resolution.  In this Newtonian problem,
the shadow has a radius equal to 2.

\vspace{0.5cm}

\section{Schwarzschild Spacetime with Gas at Rest}\label{sec:Schw}

We now carry out a general relativistic analysis for a Schwarzschild
black hole. In this section, we assume that the radiating gas is at
rest. As in the Newtonian model, we assume that the net emitted
luminosity between $r$ and infinity is $1/r$ (this is approximate, but is good enough for our purposes). To translate to the
power per unit volume $P_{\rm S}(r)$ emitted in the proper frame of the gas,
we note the following: (i) The proper length corresponding to the
coordinate interval $dr$ is $\sqrt{g_{rr}}dr$. (ii) Every unit of
energy emitted in the local frame at $r$ corresponds to
$\sqrt{-g_{tt}}$ units of energy at infinity. (iii) A time interval
$dt$ in the local frame corresponds to an interval $dt/\sqrt{-g_{tt}}$
at infinity. In view of these scalings, a natural choice for $P_{\rm S}$, as
well as the emission coefficient per unit solid angle $j_{\rm S}$, is the
following:
\begin{align}
\label{eq:Schw_emit}
P_{\rm S}(r) &= \frac{1}{|g_{tt}|\sqrt{g_{rr}} 4\pi r^4} \\
\nonumber &= \frac{1}{\left(1-\frac{2}{r}\right)^{1/2}4\pi r^4}    \\
\nonumber &= 4\pi j_{\rm S}(r), ~~ r > 2. 
\end{align}
The precise functional form and the normalization are unimportant for
what follows. The only virtue of the particular choice made here is
that it enables easy comparison with the Newtonian problem discussed
in \S3.  For instance, the net emitted luminosity in this model,
measured at infinity, is
\begin{equation}
L_{\rm emit}(r) = \int_r^\infty |g_{tt}| P_{\rm S}(r')\; 4\pi r'^2
\sqrt{g_{rr}}dr' = \frac{1}{r},
\end{equation}
the same as in the Newtonian problem. 

As before, only a fraction of the radiation emitted at any given $r$
escapes to infinity. The solid angle of the escaping rays is equal to
$2\pi(1+\cos\theta)$ for $r\geq3$ and $2\pi(1-\cos\theta)$ for $r<3$,
where $\theta$ is given by\footnote{This result can be obtained by using the fact that the critical rays that barely escape have specific angular momentum equal to that of the photon orbit: $|k_\phi/k_t|_{\rm crit} = |k_\phi/k_t|_{\rm ph} = \sqrt{27}$.} (we use the solution for which $0\leq
\theta\leq \pi/2$)
\begin{equation}
\sin\theta = \frac{3^{3/2}}{r} \left(1-\frac{2}{r}\right)^{1/2}.
\end{equation}
Counting only escaping rays, the net luminosity observed at infinity
is
\begin{equation}
L_\infty = \frac{19}{60} \approx 0.32.
\end{equation}
This is about 70\% of the luminosity at infinity in the Newtonian
problem. The reduced luminosity is because of gravitational deflection
of light rays, which causes a larger fraction of the emitted radiation
at any $r$ to be captured by the black hole.

To calculate the intensity profile $I(b)$, we need a ray-tracing code (we use the code described in
\citealt{zhu15,narayan16}). The result is shown by the green curves in
Fig.~\ref{fig:intensity_profiles}, as well as the 2D image in the
middle panel of Fig.~\ref{fig:2D}.\footnote{The results presented here are most readily verified in a fully covariant treatment of radiative transfer using a separable plasma-frame emissivity $j_{\rm S}(r) \phi_\nu$, where $\int d\nu \phi_\nu = 1$.}

Qualitatively, the image in this general relativistic model is similar
to the previous Newtonian image, with intensity increasing rapidly
with decreasing $b$, reaching a peak at a finite $b$, and then
dropping to lower values inside the peak. However, there are important
differences. First, the peak is not at $b=2$, the radius of the inner
edge of the emitting gas, but is at the radius of the photon ring $b_{\rm ph}$,
which corresponds to the lensed image of the photon orbit $r_{\rm
  ph}$. Thus the shadow is significantly larger than in the Newtonian
case.  Second, the peak intensity at $b=b_{\rm ph}$ technically goes
to infinity. This is because it is possible for rays with $b=b_{\rm
  ph}$ to make an infinite number of orbits around the black hole and
to collect an arbitrarily large intensity\footnote{This is true under the assumed optically
thin conditions. In reality, the intensity is limited to a finite maximum value because of optical depth effects.}; because of numerical limitations, the actual
computed intensity never goes to infinity.  In any case, the
divergence of the intensity at the photon ring is only logarithmic
(\citealt{gralla,johnson19}; but similar results go back many years, e.g.,
\citealt{darwin,luminet}), so the excess flux due to this divergence is
small. Finally, the central intensity $I(b=0)$ is only 25\% of that in
the Newtonian problem: $1/1536\pi^2 \approx 6.60\times 10^{-5}$.

\section{Schwarzschild Spacetime with Infalling Gas}

We now consider a more realistic model of an accreting Schwarzschild
black hole: we allow the radiating gas to move in towards the black
hole, as in a real accretion flow. In the spirit of our spherical
model, we assume that the motion is purely radial. To make a specific
choice for the radial velocity, we assume that the gas free-falls on
to the black hole from infinity. Thus we set $u_t=-1$, which
corresponds to a velocity $\beta = \sqrt{2/r}$ and Lorentz factor
$(1-2/r)^{-1/2}$ as measured by an observer at rest at radius $r$. For
simplicity, we take the radiation power and emission coefficient, as
measured in the rest frame of the infalling gas, to be the same as in
equation~(\ref{eq:Schw_emit}).

The ray-traced intensity profile $I(b)$ corresponding to this model is
shown by the red curves in Fig.~\ref{fig:intensity_profiles}, and the
2D image seen by a distant observer is shown in the right panel in
Fig.~\ref{fig:2D}. As before, the intensity increases towards smaller
$b$, reaches nearly-infinite intensity at $b=b_{\rm ph}$ and then
drops in the interior. A major new feature is that, because of
inward gas motion, radiation that is emitted isotropically in the rest
frame of the gas is beamed radially inward in the Schwarzschild
frame. The beaming is especially large at small radii.  Thus a larger
fraction of the radiation is lost into the black hole. The effect is
particularly strong in the shadow region, which is now significantly
darker than in the previous two models. For instance, $I(b=0)$ in this model is  $\approx 4.16 \times 10^{-6}$, which is a factor $\approx 16$ less than when the gas is at rest. 

\begin{figure}
\includegraphics[width=\columnwidth]{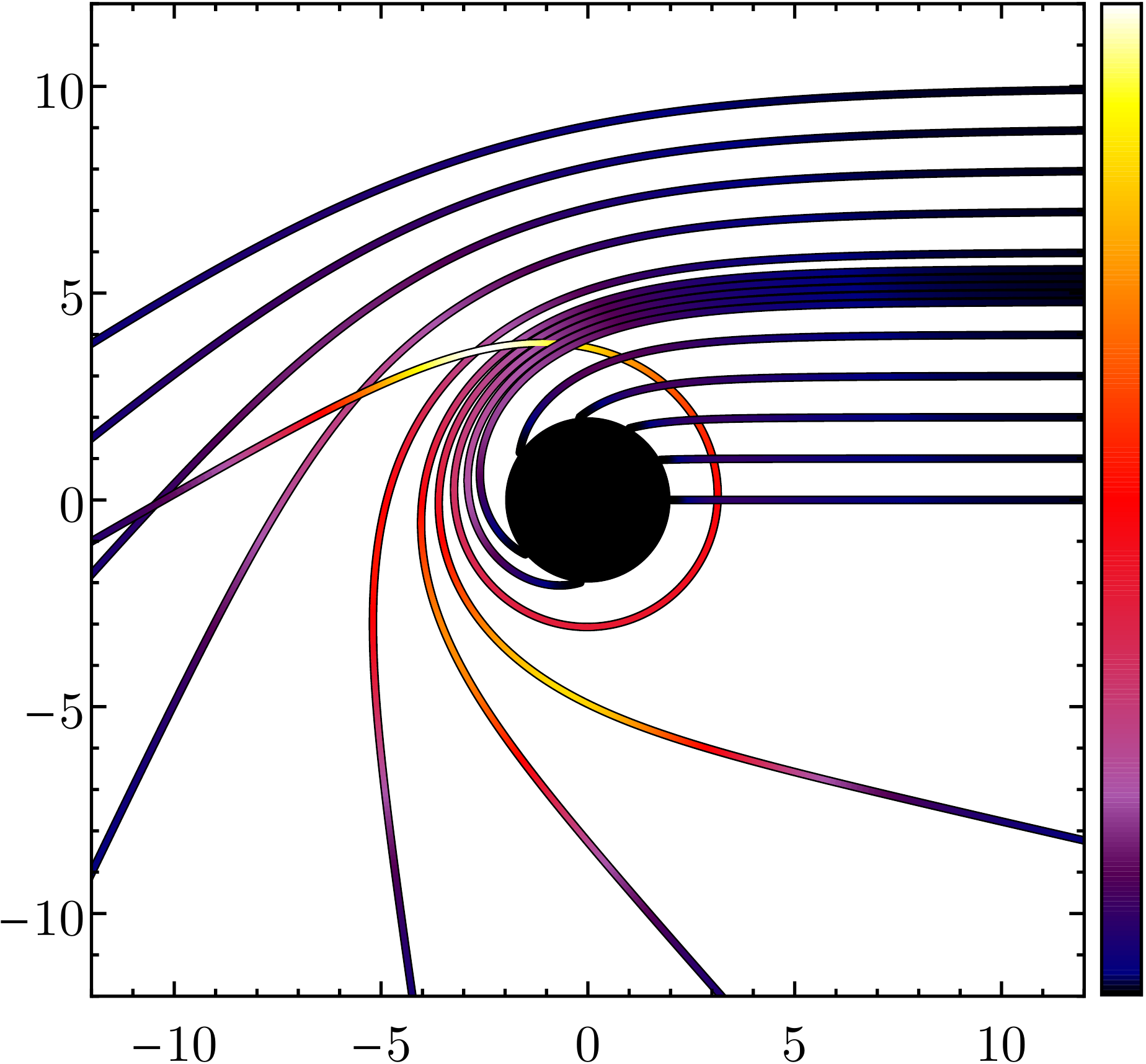}
\caption{ 
Example geodesics for a Schwarzschild black hole. Color indicates the effective emissivity along the ray for the case of radial infall, including Doppler effects. Rays are shown for impact parameters of $b = 0$ to $10$ in steps of unity, with additional rays shown from $b=4.8$ to $5.6$ in steps of 0.2 to highlight contributions near $b_{\rm ph}$. The colorscale is linear. }
\label{fig:geodesics_emissivity}
\end{figure}

Figure~\ref{fig:geodesics_emissivity} shows the emissivity contributions along representative null geodesics that travel towards an observer located to the far right of the plot. Doppler beaming causes portions of the geodesics that are moving in toward the black hole to be significantly brighter than those moving outward. Because rays with $b < b_{\rm ph}$ are always moving outward while those with $b > b_{\rm ph}$ always have both an inward and an outward segment, Doppler beaming causes a significant jump in the observed image brightness at the  edge of the shadow (rays with $b>b_{\rm ph}$ are much brighter than equivalent rays with $b < b_{\rm ph}$). This explains why the shadow in the right panel in Fig.~\ref{fig:2D} is so much deeper than that in the middle panel. For the same reason, as we show below, one obtains a deep shadow even if there is no emitting material at or near the photon orbit, $r_{\rm ph} = 3$. 

To the extent that, among the various spherical
models considered in this paper, the model with infalling gas comes closest to a
radiatively inefficient accretion flow in a low-luminosity black hole,
the shadow we compute in this section may be viewed as the best analog of the
image observed in M87 \citepalias{EHT4}.

\begin{figure*}
\includegraphics[width=0.99\textwidth]{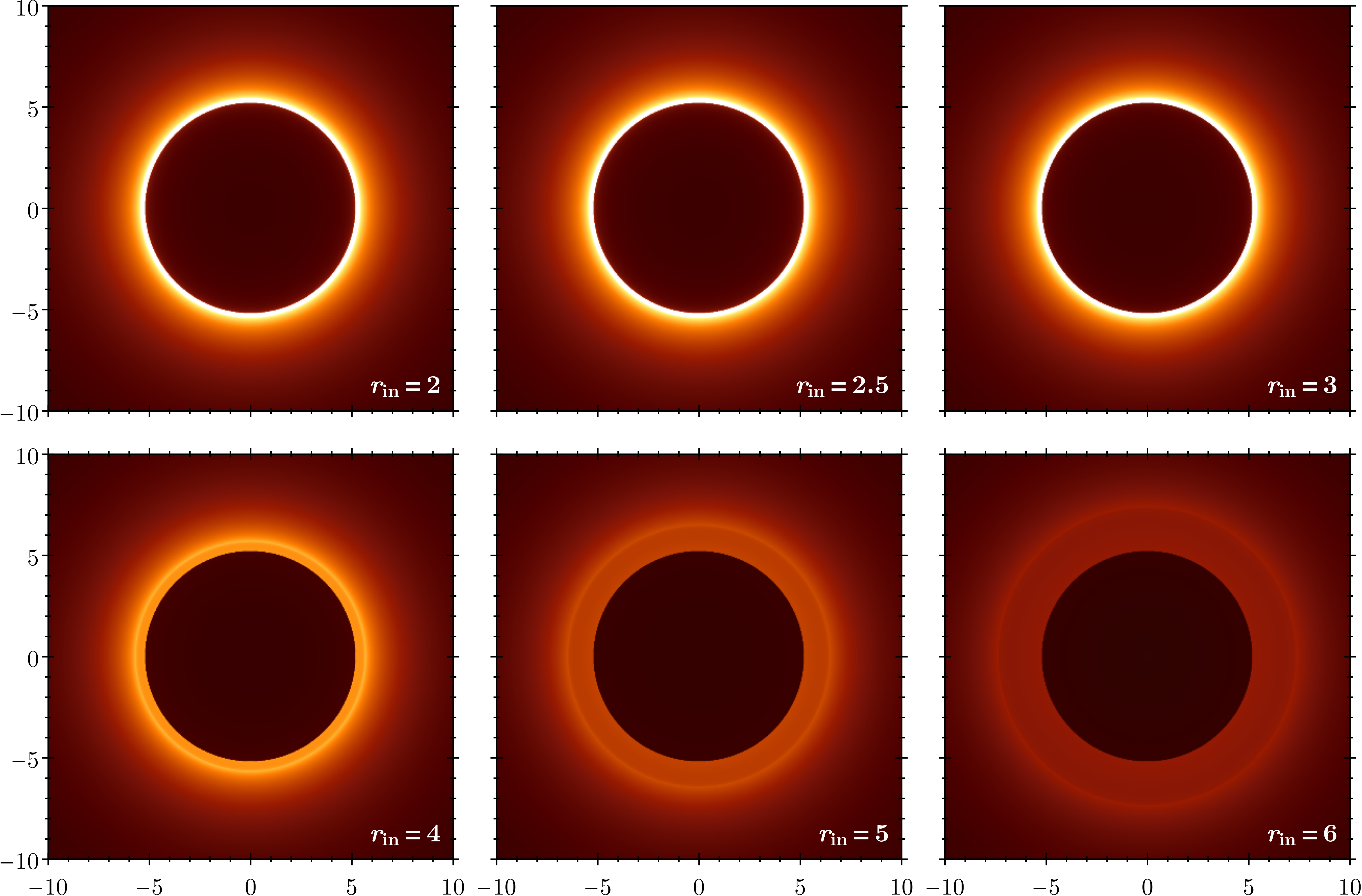}
\caption{Image seen by a distant observer for radially infalling gas
  in Schwarzschild spacetime. The emission is cut off at different
  inner radii $r_{\rm in}$ in the various panels, extending from $r_{\rm in}=2$ (Top-Left) to $r_{\rm in}=6$ (Bottom-Right).
  All panels use a linear colorscale from $0-10^{-4}$ (see Fig.~\ref{fig:2D}).}
\label{fig:rin}
\end{figure*}

\begin{figure}
\centering
\includegraphics[width=0.75\columnwidth]{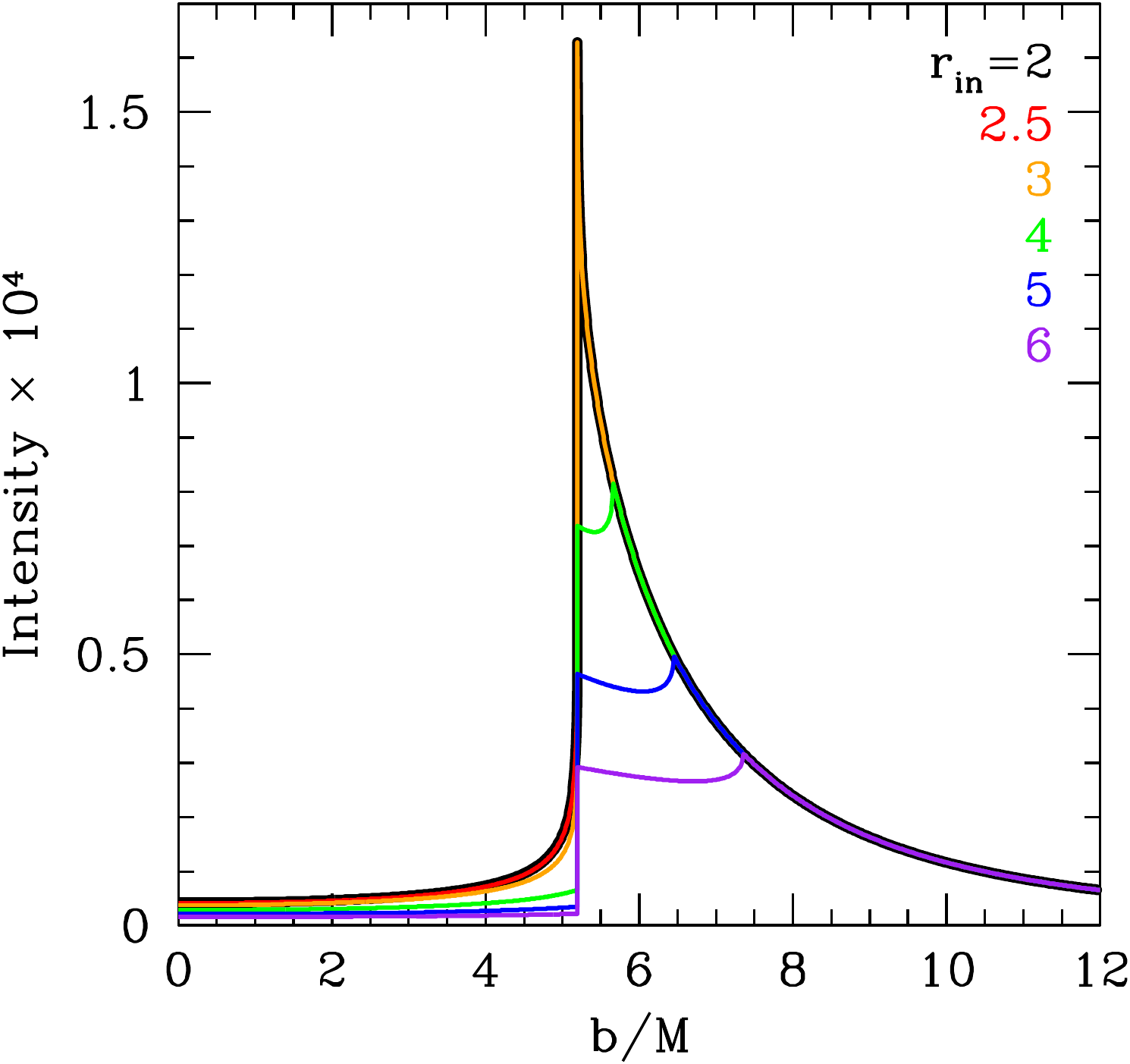}
\caption{Intensity profile $I(b)$ as a function of the
  impact parameter $b$ for the models with radially infalling gas in Schwarzschild spacetime and varying emission cutoff radius $r_{\rm in}$, as shown in Fig.~\ref{fig:rin}. In all cases, the radius of the shadow (the dark central region) is equal to $b_{\rm ph} \approx 5.2$.}
\label{fig:intensity_profiles_Schw_ff}
\end{figure}

The most striking feature of the two relativistic model images in
Fig.~\ref{fig:2D} (middle and right panels) is the fact that the
radius of the shadow is determined by the photon ring $b_{\rm ph}$, which is a
geometrical property of the spacetime. It is not determined by the inner radius of
the radiating gas, which is located at the horizon,
$r_{\rm in}=2$. Effectively, the spacetime geometry and lensing of
photon trajectories erases details of the accretion flow at small
radii, and imposes on the observed image a unique signature that reflects the properties of the
spacetime.

To emphasize this point, we show in Fig.~\ref{fig:rin} ray-traced
images for six different models, all of which have radially infalling
gas in Schwarzschild spacetime. The top left panel corresponds to the
model we just described, in which the gas radiates all the way down to
the horizon: $r_{\rm in}=2$. In the remaining five models the emission
is cutoff at progressively larger inner radii: $r_{\rm in} = 2.5$, 3,
4, 5, 6.  The first two of these models, in which the inner edge is
inside the photon orbit $r_{\rm ph}$, have images that are virtually
indistinguishable from the initial model (they have slightly darker
shadows, though it is hard to tell visually; see Fig.~\ref{fig:intensity_profiles_Schw_ff}). The remaining models, in
which $r_{\rm in} > r_{\rm ph}$, show a weak feature in the image at
the lensed location of $r_{\rm in}$. However, the most prominent
feature in all six models is the dark shadow in the middle. In each
case the outer edge of the shadow is located at the radius of the
photon ring $b_{\rm ph}$. The spherical accretion model thus provides a class of counterexamples to 
the strong claim by \citet{gralla} that the size of the shadow is very much dependent on the emission model.

The behavior of the image in the spherical model differs from that of geometrically thin disks, where the disk inner edge can have a prominent signature in the image, especially if the edge is outside the photon orbit (e.g., see \citealt{luminet}). However, the thin disk model is quite implausible for M87\footnote{\citet{gralla} give one such implausible model for M87, a thin disk terminating at the innermost stable circular orbit of a Schwarzschild black hole. This is their only 
example in which the ``effective radius'' of the image is significantly larger than the photon ring.} or other low-luminosity galactic nuclei, which are known to contain geometrically thick hot accretion flows \citep{yuan}.

\section{Discussion}

The main result of this paper is that, for a spherically symmetric
general relativistic (Schwarzschild spacetime) accretion model with
optically thin gas, the concept of a shadow in the image is
robust. The shadow is circular, and its outer edge is located at the
photon ring radius $b_{\rm ph}$, which is uniquely determined by the
spacetime metric. The outer radius of the shadow does not depend on
the details of the accretion flow. In particular, the location of the
inner edge $r_{\rm in}$ of the radiating gas has very little
effect. We demonstrate this in Figs.~\ref{fig:rin} and \ref{fig:intensity_profiles_Schw_ff}, where we consider
models from $r_{\rm in}= 2$ (the radius of the horizon) to $r_{\rm
  in}=6$ (the radius of the innermost stable circular orbit). We find
that the shadow has an identical size in all the models.

The intensity contrast between the interior of the shadow and the
emission just outside $b_{\rm ph}$ does depend on details of the
accretion. The contrast is not large when the gas is stationary, but
it is quite dramatic when the gas has a large inward radial velocity
(compare the middle and right panels in Fig.~\ref{fig:2D}). Real
accretion flows do have inward motions, and the radial velocity tends
to be large precisely at the radii of interest for shadow
formation. Hence the model with radially infalling gas is most
appropriate for comparison with the image of M87.

Real accretion flows are not spherically symmetric. The hot accretion
flow in Sagittarius A$^*$, M87, and most other galactic nuclei
consists of a geometrically thick, quasi-spherical disk \citep{yuan}, whose
characteristics lie somewhere between the pure spherical model
considered here and the geometrically thin disks considered in earlier
work (e.g., \citealt{luminet,gralla}). 

Real black holes are also expected to have nonzero spin.  We did not consider spin because the corresponding Kerr spacetime is not spherically symmetric, which complicates the model.  Analysis of emission profiles of semi-analytic models of radiatively inefficient accretion flows around spinning black holes, beginning with \citet{jaroszynski}, produce qualitatively similar results (see their Fig.~3, noticing that the brightness profiles are averaged over position angle).  More broadly, our simplified spherical model captures key features that also appear in state of the art GRMHD models \citepalias{EHT5}, whether they are spinning or not: (1) close in, the emissivity rises almost monotonically inward toward the horizon; (2) infall at $r \sim 2 G M/c^2$ leads to Doppler beaming of the emission toward the horizon, resulting in (3) low surface brightness on lines of sight that intersect the horizon: the shadow of the black hole.

An important observable signature that constrains non-spherical models is the presence of
angular momentum in the accreting gas. This causes a Doppler asymmetry
in the intensity distribution around the ring of emission outside the
shadow.  An azimuthal brightness asymmetry exceeding 2:1 is clearly seen in the observed image of M87
\citepalias{EHT4} and also in model images based on GRMHD simulations
\citepalias{EHT5}. Because M87 is viewed largely pole-on (the inclination angle
is estimated to be $17^\circ$, cf, \citealt{walker}), the brightness asymmetry from rotation in a disk is 
relatively weak. 

Another feature of real accretion that is not captured by our
spherically symmetric model is the presence of outflows.  Hot
accretion flows tend to have inward accretion in more equatorial
regions and outflows closer to the poles. Indeed, M87 has a powerful
jet, which extends out to very large distances \citep{walker}.  At
radii of interest to us ($r\sim$\,few), the outward gas velocity in
the polar regions is expected to be sub-relativistic; in fact, if
these radii are inside the ``stagnation surface,'' the gas may even be
flowing in towards the black hole (e.g., \citealt{pu}).  Regardless,
the Doppler deboosting which causes the shadow to be so dark in our
radial infall models (Fig.~\ref{fig:rin}) will be less pronounced or
even absent in the polar regions. This might cause the contrast
between the shadow and its exterior to be less pronounced \citepalias[examples
may be seen in][]{EHT5}.

Yet another practical matter is that observations never have the
infinite angular resolution needed to identify precisely the edge of
the shadow. With a finite resolution, one sees a blurred version of
the image, where the emission ring one observes outside the shadow is
a convolution of the intrinsic image intensity distribution $I(b)$ and the
point-spread function of the measurements. This often causes the peak
of the ring to shift outward to a somewhat larger radius (by about
10\%) than the photon ring radius $b_{\rm ph}$.  This effect and its uncertainties can be
quantified using simulations, and \citetalias{EHT6} included it in the estimation and error budget for the 
mass of the black hole in M87. 

Because our interest in this paper is the shadow, we focused only on
the role the photon ring plays as the location where the shadow
terminates. Another property of the photon ring is that it is a place
where the image intensity can become large. An
enhancement of the observed intensity happens whenever there is
radiating gas in the vicinity of the photon orbit $r_{\rm ph}$, and
the gas is optically thin. This effect is seen, for instance, in the
green and red curves in Fig.~\ref{fig:intensity_profiles}, in the three models with $r_{\rm in}=2$, 2.5, 3 in Fig.~\ref{fig:intensity_profiles_Schw_ff} (but not the other three models), and also in
simulation images in \citetalias{EHT5}. The integrated excess flux in the sharp
intensity peak is not very large because the divergence is only
logarithmic. However, the sharpness of the peak produces a strong and robust signature in
Fourier space, opening up the possibility of making precise
measurements of the photon ring with very long baseline interferometry
in space \citep{johnson19}.

\section*{Acknowledgements}

The authors thank Avery Broderick, Alexandru Lupsasca, Dimitrios
Psaltis, and Andrew Strominger for comments, 
and the referee for a number of useful suggestions. 
RN was supported in part by grants OISE-1743747 and AST-1816420 from the National
Science Foundation (NSF). 
MJ was supported in part by grants AST-1716536 from the NSF and GBMF-5278 from the Gordon and Betty Moore Foundation. 
CG was supported in part by grants AST-1716327 and OISE-1743747 from the NSF. 
This work was carried out at the Black Hole Initiative, Harvard University, which is
funded by grants from the John Templeton Foundation and the Gordon and
Betty Moore Foundation to Harvard University.

\end{document}